\documentclass{PoS}
\usepackage{graphicx}
\usepackage{amsmath}
\usepackage{amssymb}
\usepackage{amsthm}
\usepackage[section]{placeins}
\title{Tensor renormalization group study of the 2d O(3) model}

\ShortTitle{Tensor renormalization group study of the 2d O(3) model}

\author{\speaker{Judah Unmuth-Yockey}%
        \\
        University of Iowa\\
        E-mail: \email{judah-unmuth-yockey@uiowa.edu}}

\author{Yannick Meurice\\
       University of Iowa\\
       E-mail: \email{yannick-meurice@uiowa.edu}}

\author{James Osborn\\
       Argonne National Laboratory\\
       E-mail: \email{osborn@alcf.anl.gov}}

\author{Haiyuan Zou\\
       University of Pittsburgh\\
       E-mail: \email{haz57@pitt.edu}}

\abstract{We present our progress on a study of the $O(3)$ model in
two-dimensions using the Tensor Renormalization Group method.  We first construct
the theory in terms of tensors, and show how to construct $n$-point correlation functions.  We then give results for thermodynamic quantities at finite and
infinite volume, as well as 2-point correlation function data.  We discuss some
of the advantages and challenges of tensor renormalization and future directions in which to work.}

\FullConference{The 32nd International Symposium on Lattice Field Theory\\
                 23-28 June, 2014\\
                 Columbia University New York, NY}

\begin{document}
    \section{The Model \& Tensor Renormalization}
        In the following we consider the $O(3)$ non-linear sigma model with
        Hamiltonian
        \begin{equation}
            H = -\sum_{\langle ij \rangle} \vec{S}_{i} \cdot \vec{S}_{j}
        \end{equation}
        where $\vec{S}$ is a three-component vector, and $\langle ij \rangle$
        a sum over nearest-neighbor pairs on a two-dimensional square lattice.
        The partition function is
        \begin{equation}
            Z(\beta) = \prod_{x} \int d\Omega(x) e^{-\beta H}.
        \end{equation}
        The sum is over spin configurations on a two-dimensional lattice, and $d\Omega$ is the unit spherical area element.
        This model is important for a number of reasons.  Firstly, it has a non-Abelian global symmetry.  Second, the model has no spontaneous symmetry
        breaking (i.e. $\langle \vec{S} \rangle = 0$ for all $\beta$), unlike the Ising model in two dimensions.
        Thirdly, this model is known to be asymptotically free for large $\beta$.
        Lastly, the model also possesses instantons which are labeled by an integer
        topological index.
    
\begin{figure}
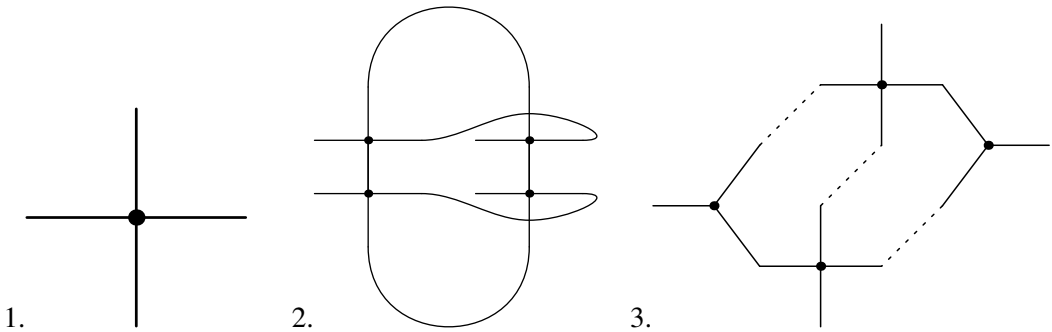

        \centering
            1.\includegraphics[width=0.20\textwidth]{singletensor.3} \quad
            2.\includegraphics[width=0.25\textwidth]{mtensor.3}\quad
            3.\includegraphics[width=0.35\textwidth]{tensorupdate.3}
            \caption{The steps involved in TRG blocking with truncation.}
            \label{trgprocess}
        \end{figure}
\begin{figure}
        \centering
            \includegraphics[width=0.3\textwidth]{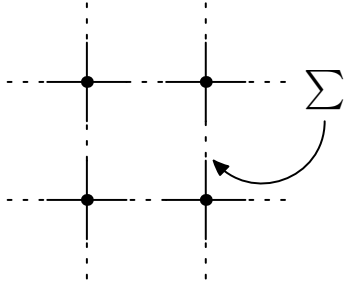}
            \caption{A graphical representation of building up the lattice by
            contracting tensors.}
            \label{trglattice}
        \end{figure}

        The Tensor Renormalization Group (TRG) \cite{levin} is a renormalization group method developed for classical statistical models.
        TRG consists of reformulating a classical partition function in terms of
        local tensors.
        The techniques used here are similar to \cite{xie}, which hinge on
        using Higher Order Singular Value Decomposition (HOSVD), giving the renormalization
        group technique the name Higher Order Tensor Renormalization Group (HOTRG).
        The methodology is as follows. First one must form the initial tensor, or devise a tensor formulation for one's model.  Second, one performs a
        blocking step.  A blocking step includes contracting two tensors together, then
        using the HOSVD to pick out the most relevant directions in renormalization group product space.  Once we pick out the relevant directions in renormalization group
         space, we project
        out those states to form a new tensor. This process is illustrated
        in Figure \ref{trgprocess}.
        The tensor formulation allows one to decouple the lattice at the location of the local constraint, in our case, the sites of the lattice.
        The lattice is then rebuilt by piecing these tensors together geometrically in the shape of the lattice.  The idea of contracting the tensors to build up a lattice is shown in
        Figure \ref{trglattice}.

    \section{Tensor Formulation for $O(3)$}
\begin{figure}
        \centering
            \includegraphics[width=0.6\textwidth]{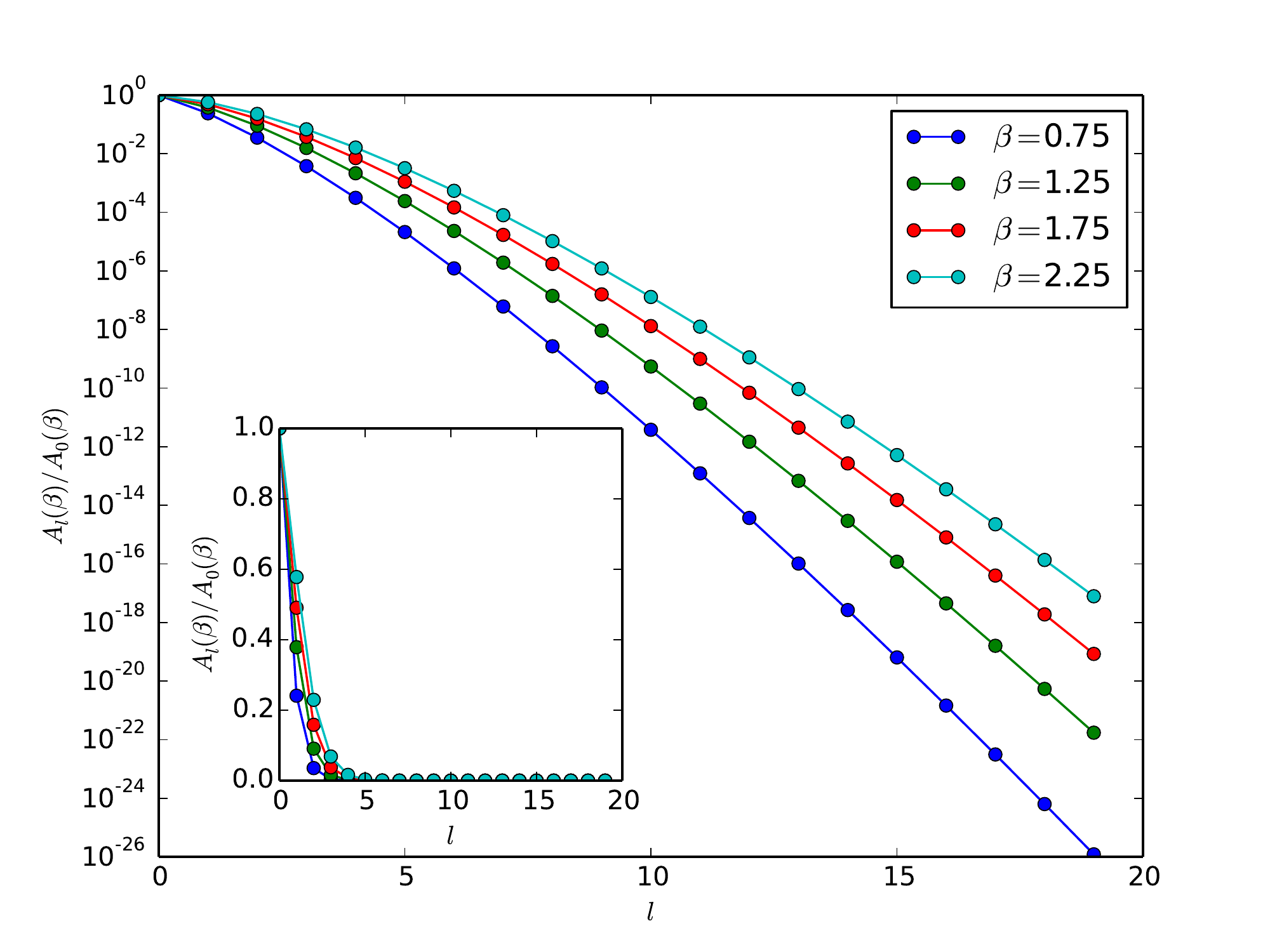}
            \caption{The relative size of the weights as a function of total
            angular momentum number $l$}
            \label{coeffalloff}
        \end{figure}

\begin{figure}
        \centering
            \includegraphics[width=0.4\textwidth]{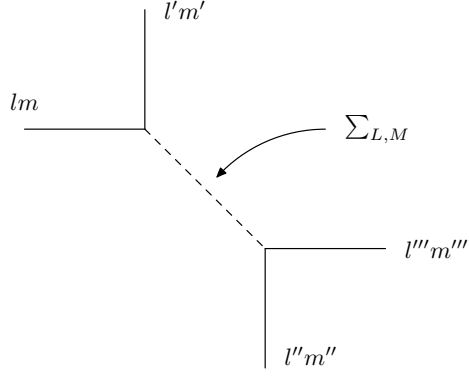}
            \caption{The local tensor for the $O(3)$ model.}
            \label{o3tensor}
        \end{figure}

        A simple approach to formulating the $O(3)$ model in terms of tensors
        is through harmonic analysis.
        $O(3)$ has two quantum numbers: $l$ and $m$.  However the expansion
        coefficients, when expanding in terms of Legendre polynomials, only depend on $l$.  To proceed we expand the Boltzmann
        weight in terms of Legendre polynomials
        \begin{equation}
            \exp[\beta \cos\gamma_{ij}] = \sum_{l}A_{l}(\beta) P_{l}(\cos\gamma_{ij}).
        \end{equation}
        The relative size of the coefficients
        can be seen in Figure \ref{coeffalloff}.  Now 
        \begin{equation}
            P_{l}(\cos\gamma_{ij}) = \frac{4 \pi}{2l+1}\sum_{m} Y_{lm}(\theta_{i},\phi_{i}) 
        Y^{*}_{lm}(\theta_{j},\phi_{j}).
        \end{equation}
        This allows us to factorize the initial contribution from the links.  With the angular dependence decoupled, angular integration can now take place, and the theory can be written in terms of integer-valued fields
        $l$, and $m$.
        The spherical harmonics are associated with pairs of sites, or links.
        Since there are four impinging links per site on the lattice, the angular integration is built out of four spherical harmonics.  This integral is responsible for the constraint at a site and has the form
        \begin{align}
            &\int d\Omega \; Y_{lm}^{*} Y_{l'm'}^{*} Y_{l''m''} Y_{l'''m'''}(\Omega) \\
            &= \frac{\sqrt{(2l+1)(2l'+1)(2l''+1)(2l'''+1)}}{4 \pi}
             \sum_{L = |l-l'|}^{l+l'} \frac{1}{(2L+1)} \sum_{M = -L}^{L} C_{lml'm'}^{LM}
        C_{l0l'0}^{L0} C_{l''m''l'''m'''}^{LM} C_{l''0l'''0}^{L0}.
        \end{align}
        This constraint enforces the triangle in-equalities between pairs of
        external legs of the tensor and intermediate angular momentum quantum
        numbers. The intermediate
        sum over $L$ and $M$ picks out irreducible representations of the angular momenta.  A picture of how one might interpret the $O(3)$ tensor is shown
        in Figure \ref{o3tensor}.  The tensor can be written down explicitly as
        \begin{equation}
            T_{(l,m)(l'm')(l''m'')(l'''m''')}(\beta, x) =
            \sqrt{A_{l}A_{l'}A_{l''}A_{l'''}(\beta)}
            \mathcal{C}[(l,m)(l'm')(l''m'')(l'''m'''), x]
        \end{equation}
        with $\mathcal{C}$ the constraint from before, and $x$ the location on the
        lattice.
        This is in contrast with the Abelian constraint, which simply
        demands a conservation of the integer-valued fields associated
        with the links at their common site.
        From this tensor the entire partition function and the free energy can be built up and thermodynamical quantities calculated.

    \section{$n$-point Correlations}
        $n$-point correlations can be realized on the lattice by inserting
        spin vectors at particular sites.  This comes from the typical expression
        for correlation functions
        \begin{equation}
            \langle \vec{S}_1 \vec{S}_2 \ldots \vec{S}_n \rangle = 
            \prod_{x} \int d\Omega(x) (\vec{S}_1 \vec{S}_2 \ldots \vec{S}_n) e^{-\beta H}.
        \end{equation}
        These additional spin vectors lead to a modified constraint at each site of insertion.  This comes
        about by expanding the Boltzmann weight as usual, however there are additional degrees of freedom to integrate due to the modified integrand.
        The integrand can be made more suitable for integration by using the
        spherical basis,
        \begin{equation}
            S_{\mu} = \sqrt{\frac{4\pi}{3}}|S|Y_{1\mu}(\theta,\phi),
        \end{equation}
        as a representation for the spin vectors.
        The angular integral that needs to be preformed at each site
        of insertion now has the form
        \begin{align}
        &\int d\Omega \; Y_{l_1 m_1} Y_{l_2 m_2}
        Y^{*}_{l_3 m_3} Y^{*}_{l_4 m_4} Y_{1 m}(\Omega) \\
        & = \frac{\sqrt{(2l_1 + 1)(2l_2 + 1)(2l_3 + 1)(2l_4 + 1)}}{4 \pi}
        \sum_{LL'MM'} C^{L0}_{l_1 0 l_2 0}
        C^{LM}_{l_1 m_1 l_2 m_2} C^{L'0}_{l_3 0 l_4 0}
        C^{L'M'}_{l_3 m_3 l_4 m_4} \\
        & \times \frac{1}{(2L' + 1)} C^{L'0}_{L 0 1 0}
        C^{L'M'}_{L M 1 m}.
        \end{align}
        This leads to a tensor whose elements are shifted, or an ``impure'' tensor.
        \begin{figure}
        \centering
            \includegraphics[width=0.5\textwidth]{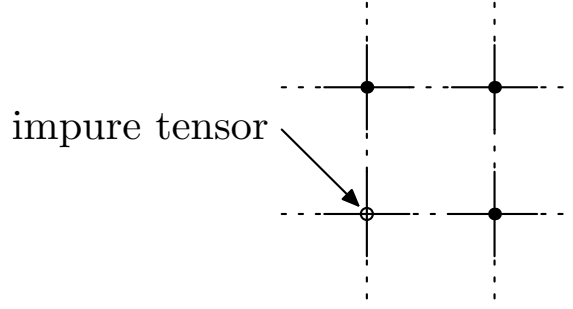}
            \caption{The placement of an impure tensor in the lattice, and contracting using TRG.  This represents inserting an additional spin
            at this lattice location.}
            \label{implat}
        \end{figure}
        Average values are computed by contracting these impure tensors with pure
        tensors using the same algorithmics.  The number of impure tensors used
        during the contraction pattern (and when applicable, their relative
        separation) determines which $n$-point function one is computing, and what
        modes one is considering.  For $2$-point functions it is necessary to have a
        source and a sink, i.e. two impure tensors in the lattice (in the absence of a magnetic field).  A picture of
        the insertion of a spin into the lattice using an impure tensor is shown
        in Figure \ref{implat}.

    \section{Results}
\begin{figure}
        \centering
            \includegraphics[width=0.6\textwidth]{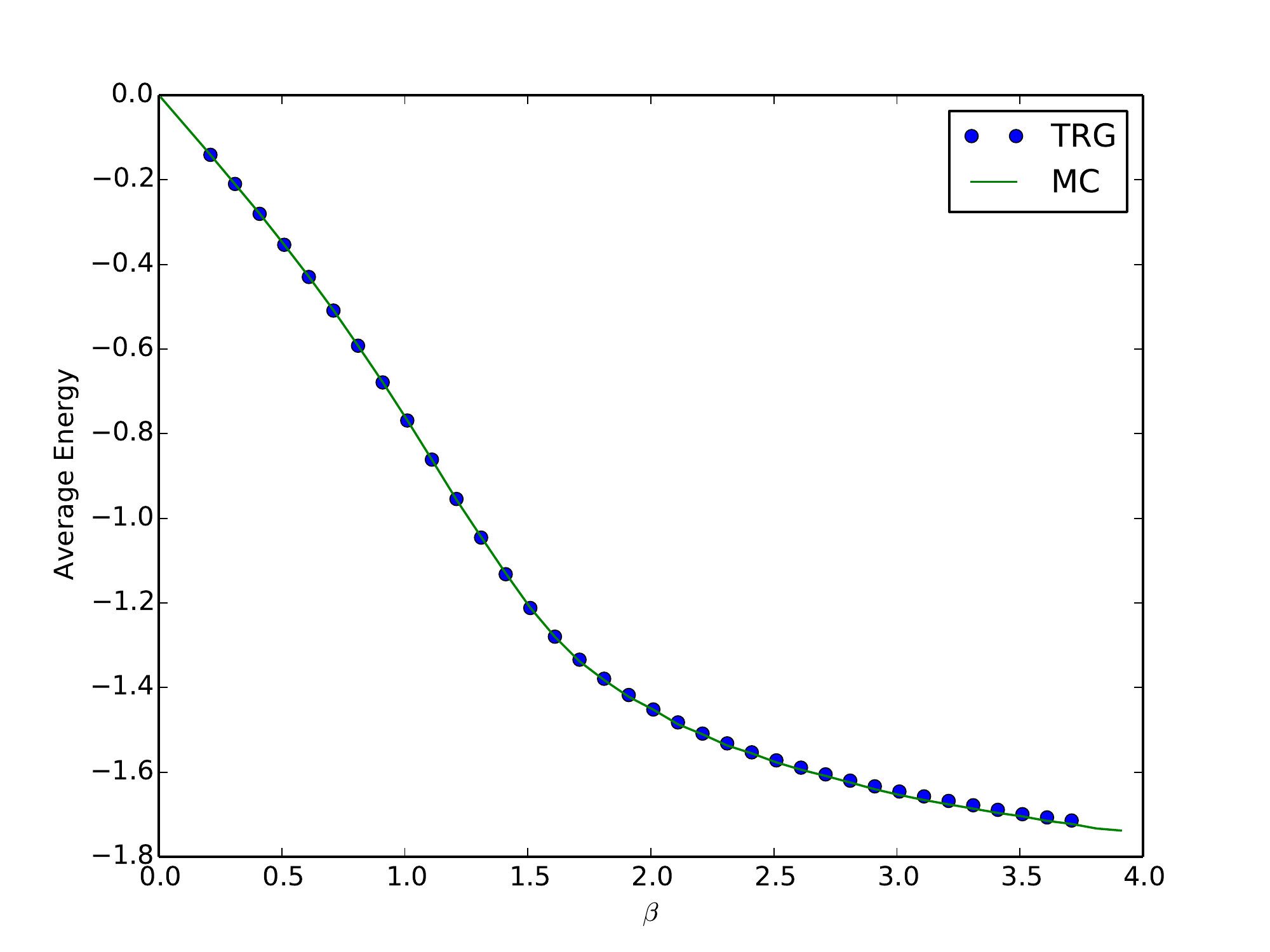}
            \caption{The average energy on a $32\times 32$ lattice using Monte
            Carlo and TRG.}
            \label{mctrgcompare}
        \end{figure}

\begin{figure}
            \includegraphics[width=0.5\textwidth]{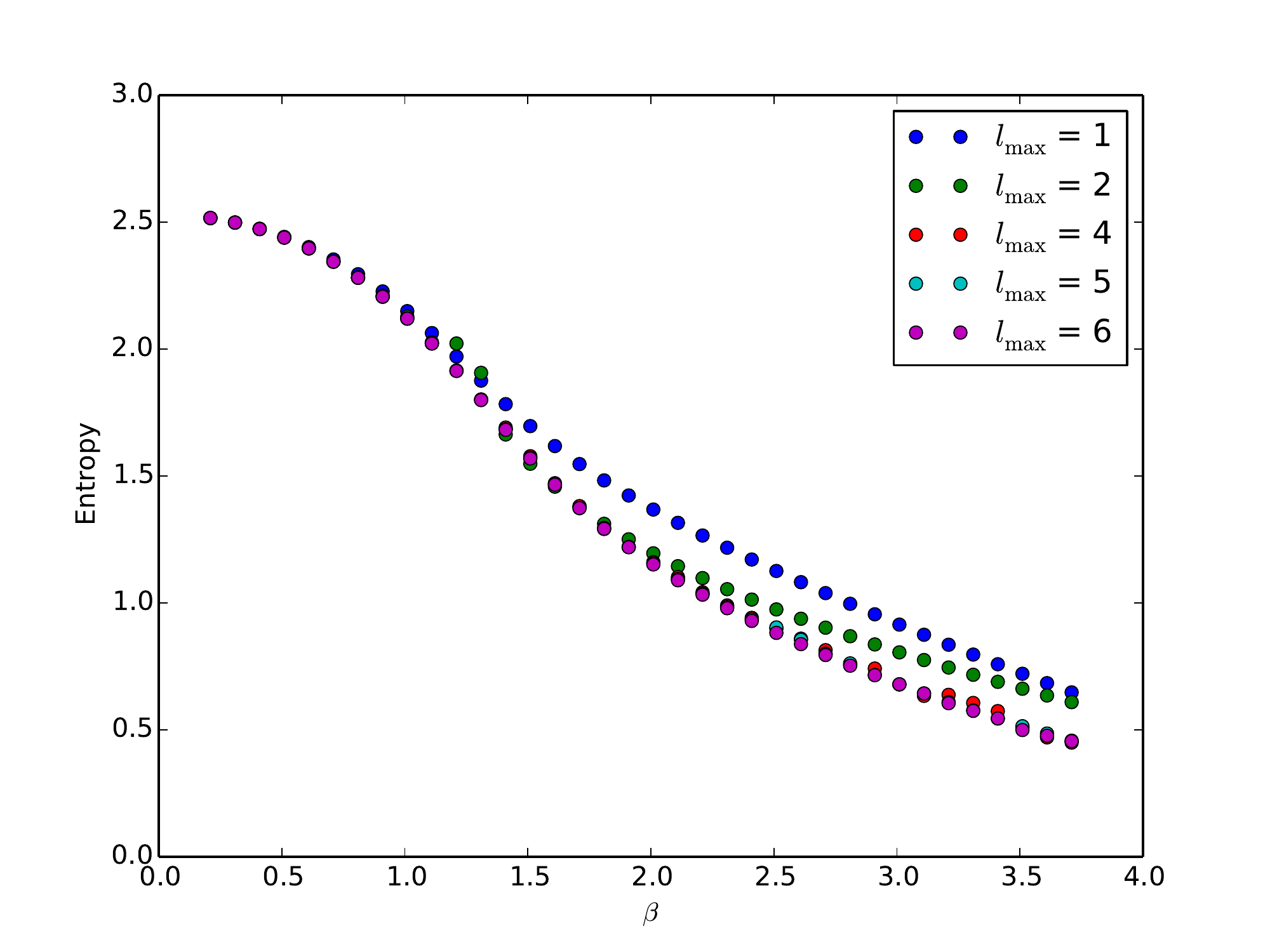}
            \includegraphics[width=0.5\textwidth]{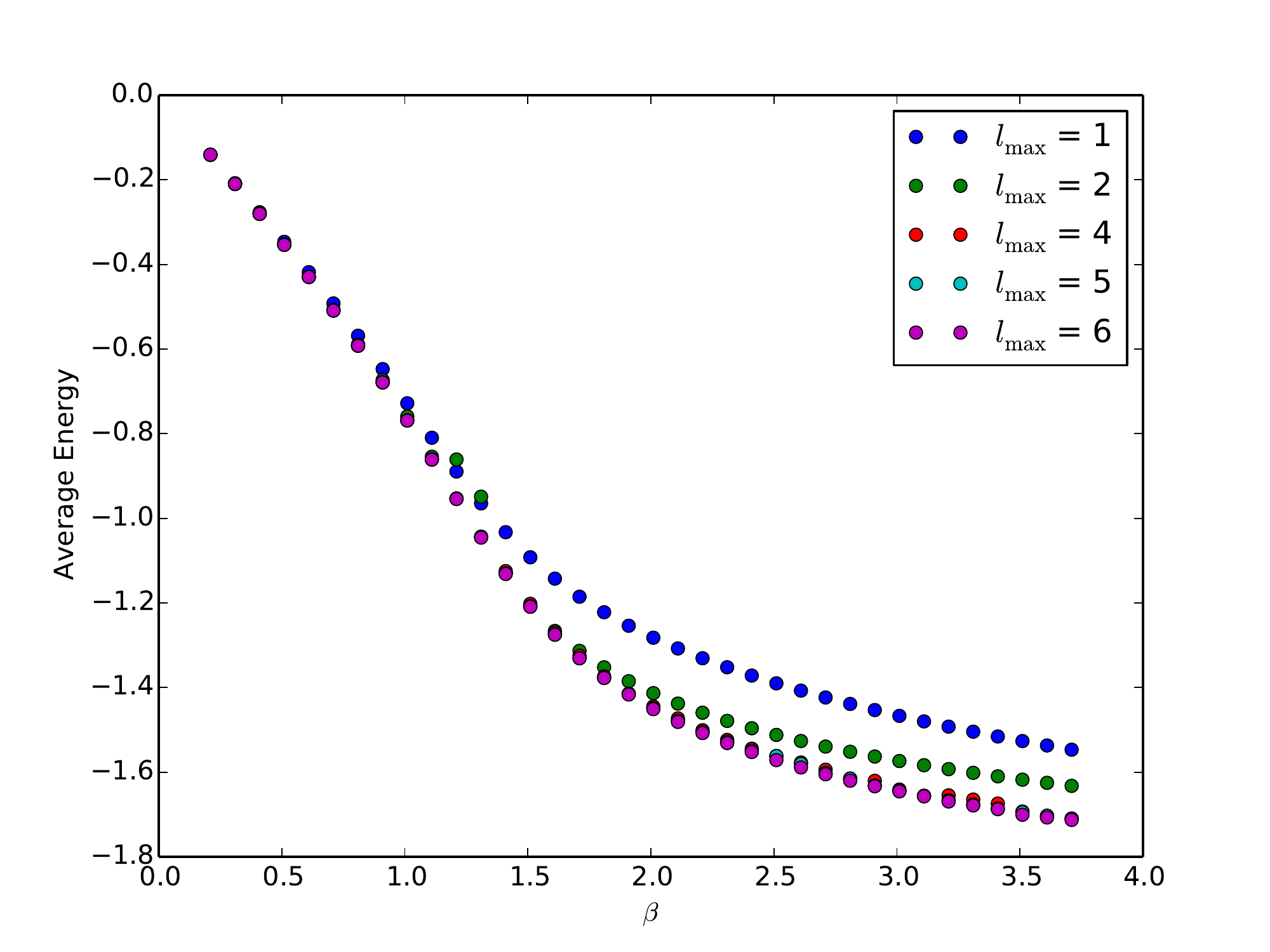}
        \caption{The average energy and entropy for the $O(3)$ model
        at approximately infinite volume.}
            \label{energyentropy}
        \end{figure}

\begin{figure}
        \includegraphics[width=0.5\textwidth]{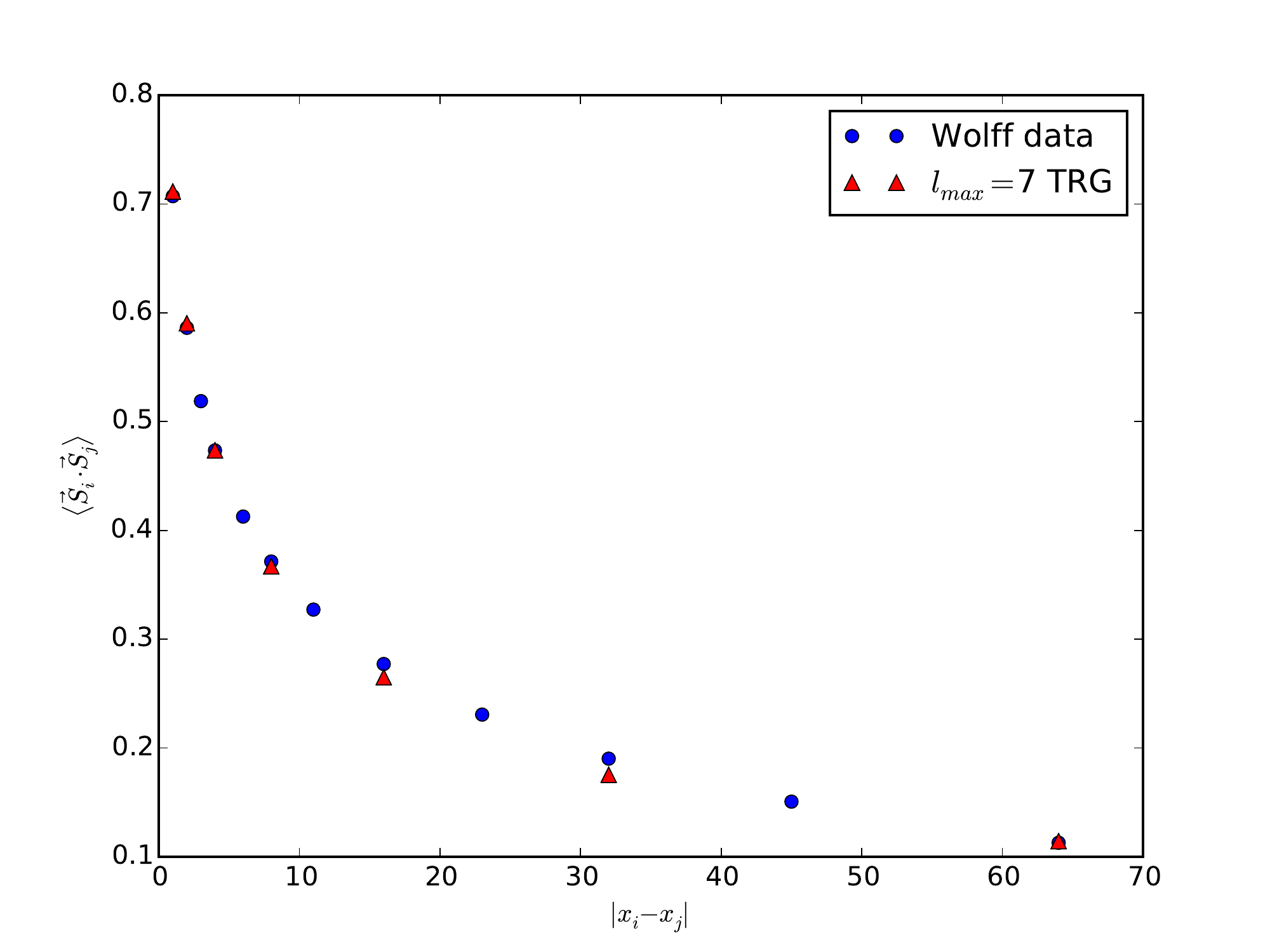}
        \includegraphics[width=0.5\textwidth]{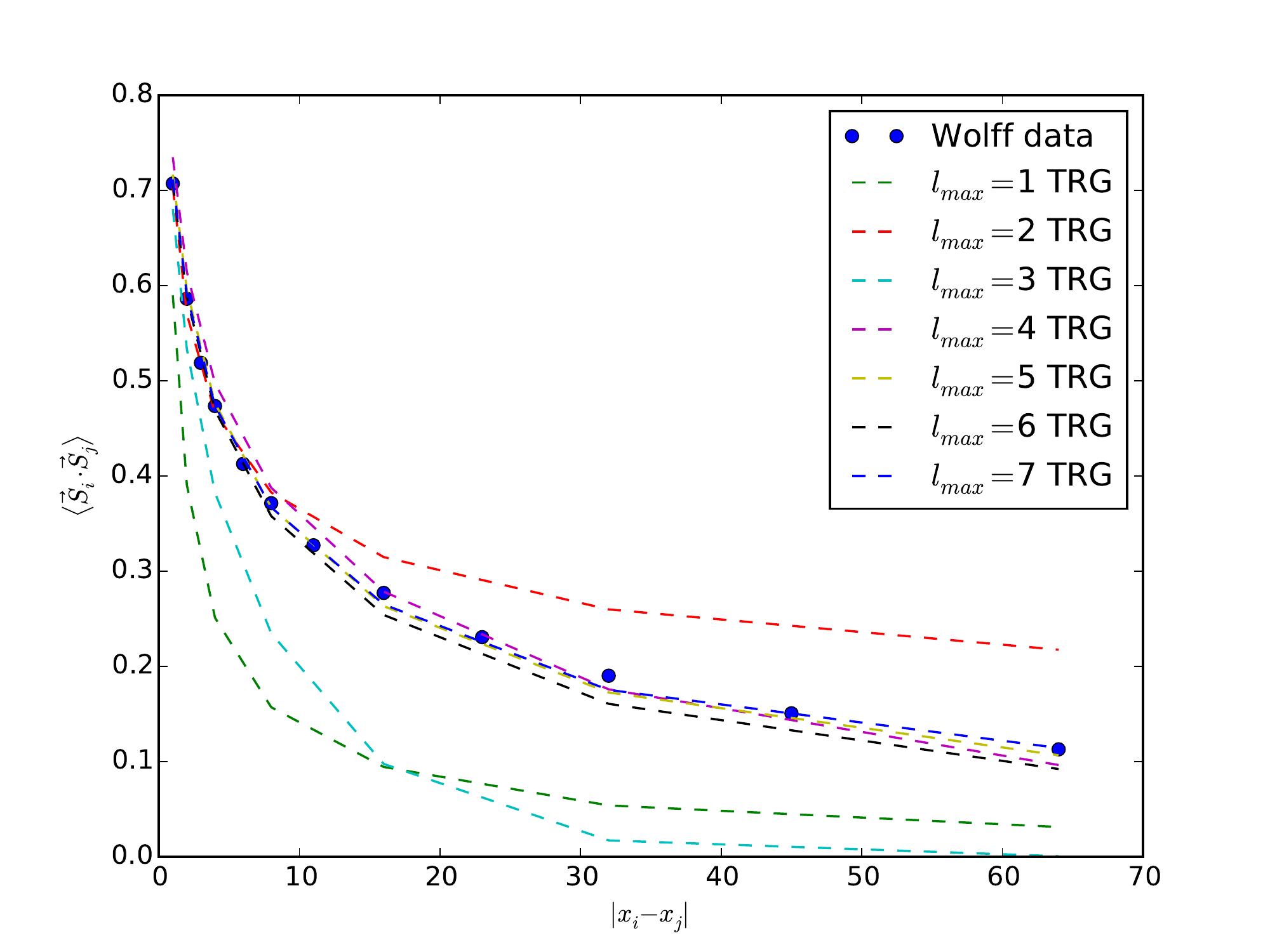}
    \caption{A comparison to the literature using TRG. The left plot has
    the TRG data at a large number of states, and the right plot shows the progression
    in the number of states compared to Table 2 in Ref. \cite{wolff2}.}
        \label{wolffcompare}
    \end{figure}

        Here we report some of the results obtained during this study.  While it has
        been shown that TRG is able to reproduce numerical data comparable to Monte
        Carlo simulations for the $O(2)$ model \cite{yu}, it was nevertheless important to cross check observables for $O(3)$ since it has not been done using TRG.  In Figure \ref{mctrgcompare} we show a comparison between Monte Carlo and TRG for the average energy at finite volume.
        It was then best to use TRG
        to explore regions where Monte Carlo has difficulty, namely, the infinite volume limit.  In Figure \ref{energyentropy} we see results for the infinite volume
        limit for average energy and entropy.  The number of states kept during iterations is given in terms of the total angular momentum number $l_{\text{max}}$.  For
        $O(3)$ the number of states, $D$, in terms of $l_{\text{max}}$ is given by $D = (l_{\text{max}}+1)^2$.
        Finally, in Figure \ref{wolffcompare}
        we see data from TRG for the 2-point correlation function and the comparison to Table 2 in Ref. \cite{wolff2}.  

    \section{Conclusion and Future Work}
        Tensor renormalization is a powerful method in 2D for extracting
        infinite volume thermodynamics.  However, it appears to give inaccurate results
        for $2$-point correlation functions so far.  TRG appears insensitive to the sign problem \cite{denbleyker}
        and computing in the infinite volume limit is as easy as computing at finite volume, since if one can compute a single block-spin, one can compute any number of them.
        It's also possible to formulate many popular models in
        terms of tensors, and a tensor formulation comes with a convenient 
        graphical representation \cite{liu}.

        Another point of interest is the fact that particular tensor elements in
        the $O(3)$ tensor formulation are negative.  It is unclear at this point if
        this formulation has negative weights for configurations.  Alternatively, there are reformulations of the $O(3)$ model where all the weights are
        positive definite \cite{wolff2010}.  It would be interesting to attempt
        other tensor formulations where one can be sure that the weights are positive.
        
        Understanding
        the systematic improvement of TRG and how the number of states kept during
        iterations improves accuracy is also a point of interest.  While ideally if one could keep all the configurations of the system the results would be exact, one can only keep a finite amount, and it becomes important to 
        understand the optimal way to keep higher numbers of states.
        In terms of scaling, TRG scales as $D^{7}$ in CPU time in two-dimensions, and the memory scales as $D^{4}$. Parallelization of tensor contractions could help the $D^{7}$ scaling.


\vskip5pt
\noindent
{\bf Acknowledgements}
       
This research was supported in part by the Department of Energy under Award Number DOE grant  DE-SC0010114. In addition, this research used resources of the Argonne Leadership Computing Facility, which is a DOE Office of Science User Facility supported under Contract DE-AC02-06CH11357.

\end{document}